\documentclass{article}

\usepackage{arxiv}

\usepackage[utf8]{inputenc} 
\usepackage[T1]{fontenc}    
\usepackage{hyperref}       
\usepackage{url}            
\usepackage{booktabs}       
\usepackage{amsfonts}       
\usepackage{nicefrac}       
\usepackage{microtype}      
\usepackage{graphicx}

\usepackage{xcolor}  
\usepackage{soul}  

\title{Time-Lagged t-Distributed Stochastic Neighbor Embedding
(t-SNE) of Molecular Simulation Trajectories}

\author{
  Vojtěch Spiwok \\
  Department of Biochemistry and Microbiology\\
  University of Chemistry and Technology, Prague\\
  Technická 5, Prague 6, 166 28, Czech Republic \\
  \texttt{spiwokv@vscht.cz} \\
   \And
  Pavel Kříž \\
  Department of Mathematics\\
  University of Chemistry and Technology, Prague\\
  Technická 5, Prague 6, 166 28, Czech Republic \\
  \texttt{krizp@vscht.cz} \\
}

\begin{document}
\maketitle

\begin{abstract}
Molecular simulation trajectories represent high-dimensional data.
Such data can be visualized by methods of dimensionality reduction.
Non-linear dimensionality reduction methods are likely to be more
efficient than linear ones due to the fact that motions of atoms
are non-linear. Here we test a popular non-linear t-distributed
stochastic neighbor embedding (t-SNE) method on analysis
of trajectories of alanine dipeptide dynamics and Trp-cage
folding and unfolding. Furthermore, we introduced a time-lagged
variant of t-SNE in order to focus on slow motions in the molecular
system. This time-lagged t-SNE efficiently visualizes slow dynamics
of molecular systems.
\end{abstract}


\section{Introduction}
The main goal of molecular simulations is identification of
key states of studied systems and building of thermodynamic
and kinetic models of transitions between these states.
Identification of key states is often based on some
numerical descriptors known as collective variables.
Distance between two atoms can be seen as one of
the simplest collective variables. It can be used,
for example, to distinguish between the bound and
unbound state in a simulation of protein-ligand interaction.
For some more complex processes it is necessary to use more
complex collective variables.

Collective variables are in fact dimensionality reduction
methods because they represent high dimensional structure
of a molecular system using few numerical descriptors.
It is therefore no surprise that general linear and
nonlinear dimensionality reduction methods have been
applied on molecular simulation trajectories. Namely,
principal component analysis \cite{ed},
diffusion maps \cite{dm},
sketch map \cite{sm1,sm2},
Isomap \cite{iso1,iso2},
autoencoders \cite{chen} or
t-SNE \cite{tsne1,tsne2} have been tested
in analysis of trajectories,
data compression or sampling enhancement.

Advantage of nonlinear dimensionality reduction methods is their
ability to describe more variance in data compared to linear methods
with the same number of dimensions. This is especially true for
t-distributed Stochastic Neighbour
Embedding (t-SNE) \cite{tsne3}.
This method became highly popular in many
fields, including data science, bioinformatics and computational
linguistics.

There are two features of t-SNE that contributed to its success.
First, t-SNE converts high-dimensional points into low-dimensional
points in a way to reproduce their proximity rather than distances.
For example, for a bioinformatician analyzing genomic data to
develop genomics-based diagnosis it is important that samples with
the same diagnosis are close to each other after dimensionality
reduction. It is unimportant how distant are samples with different
diagnosis, provided that they are distant enough. Second, t-SNE
unifies density of low-dimensional points in the output space. This
feature, which can be controlled by a parameter called perplexity,
makes visual representation of points more effective.

Disadvantage of application of general dimensionality reduction
methods on molecular simulation trajectories is that
these methods pick the most intensive (in terms of changes of atomic
coordinates) motions in the system. However, such motions are often
not interesting. Instead, for building of thermodynamic and kinetic
models or to enhance sampling it is useful to extract slowest (i.e.
energetically most demanding) motion. This can be done by
Time-lagged Independent Component Analysis (TICA)
\cite{tica1,tica2,tica3}.
TICA extracts
slowest motions in the molecular system because it correlates
the state of system with the state of the same system after a short
delay (lag). This lag can be controlled.

Here we attempt to join advantages of t-SNE and TICA into a single
method of time-lagged t-SNE. The method was tested on two molecular
trajectories -- on 200 ns simulation of alanine dipeptide and
208.8 $\mu$s simulation of Trp-cage mini-protein folding and
unfolding (trajectory kindly provided by
DE Shaw Research)\cite{shaw}.

\section{Methods}
Time-laged t-SNE is inspired by implementation of TICA using the AMUSE
algorithm \cite{amuse}. We start with atomic coordinates $\mathbf{X}(t)$
recorded over time $t$. First, coordinates are
superimposed to a reference coordinates of the system to eliminate
translational and rotational motions. After that, time-averaged
coordinates are subtracted, leading to coordinates
$\mathbf{X'}(t)$. Next, a covariance matrix is calculated as:

\begin{equation}
C^{\mathbf{X'}}_{ij} = \langle X'_{i}(t)X'_{j}(t) \rangle,
\end{equation}

where $i$ and $j$ are indexes of atomic coordinates and
$\langle \rangle$ denotes time-averaging. Next, covariance matrix is decomposed
to eigenvalues $\mathbf{\lambda^{X'}}$ (the square matrix with eigenvalues
on diagonal and zeros elsewhere) and
eigenvectors $\mathbf{W^{X'}}$ (the matrix with eigenvectors as columns):

\begin{equation}
\mathbf{C^{X'}} \mathbf{W^{\mathbf{X'}}} =
\mathbf{W^{\mathbf{X'}}} \mathbf{\lambda^{\mathbf{X'}}} .
\end{equation}

Coordinates $\mathbf{X'}(t)$ are transformed onto
principal components
and normalized by roots of eigenvalues (space-whitening the signal):

\begin{equation}
\mathbf{Y}(t) = \mathbf{(\lambda^{\mathbf{X'}})}^{-1/2}
( (\mathbf{W^{\mathbf{X'}}})^{T} \mathbf{X'}(t) ) .
\end{equation}

A time-lagged covariance matrix is calculated as:

\begin{equation}
C^{\mathbf{Y}}_{ij} = \langle Y_{i}(t)Y_{j}(t+\tau) \rangle,
\end{equation}

where $\tau$ is an adjustable time lag. Because the matrix $C$
is non-symmetric it must be symmetrized as:

\begin{equation}
\mathbf{C}^{\mathbf{Y}}_{sym} = 1/2 (\mathbf{C^{Y}} + \mathbf{(C^{Y})}^T) .
\end{equation}

Next, this symmetric matrix is decomposed
to eigenvalues $\mathbf{\lambda^{Y}}$ and
eigenvectors $\mathbf{W^{Y}}$:

\begin{equation}
\mathbf{C}^{\textbf{Y}}_{sym} \mathbf{W^{\mathbf{Y}}} =
\mathbf{W^{\mathbf{Y}}} \mathbf{\lambda^{\mathbf{Y}}} .
\end{equation}

Finally, $\mathbf{Y(t)}$ are transformed onto
principal components and expanded by eigenvalues:

\begin{equation}
\mathbf{Z} =  \lambda^{\mathbf{Y}}((\mathbf{W^{\mathbf{Y}}})^{T} \mathbf{Y} ).
\end{equation}

This step expands distances in directions with highest autocorrelations,
which represent directions with slow motions.

It is possible to use certain number of eigenvectors with
highest eigenvalues instead of all eigenvectors.
t-SNE can be applied on distances between simulation snapshots
calculated in the space of $\mathbf{Z}$ as:

\begin{equation}
D_{t,t'}=\|\mathbf{Z}(t)-\mathbf{Z}(t')\|.
\end{equation}

All analyses were done by programs written in Python with
MDtraj \cite{mdtraj}, 
PyEMMA \cite{pyemma},
numpy \cite{numpy} and
scikit-learn \cite{scikit}
libraries. It is available
at GitHub (\url{https://github.com/spiwokv/tltsne}) and using PyPI.

The trajectory of alanine dipeptide was obtained by unbiased
200 ns molecular dynamics simulation of a system containing 
alanine dipeptide and 874 water molecules in Gromacs \cite{gmx}.
It was modelled by Amber99SB-ILDN force field \cite{ildn}.
Electrostatic interactions
were treated by particle-mesh Ewald method \cite{pme}.
Temperature was kept constant by V-rescale thermostat \cite{bussi}.

The trajectory of Trp-cage folding and unfolding was kindly
provided by DE Shaw Research.

\section{Results}
The method was tested on two molecular systems -- on alanine
dipeptide and Trp-cage. In order to test time-lagged t-SNE
we compare time-lagged t-SNE with standard t-SNE and TICA.

\subsection{Alanine Dipeptide}
Time-lagged t-SNE was first applied on a trajectory of
alanine dipeptide without water and hydrogen atoms. It is
important to remove hydrogen atoms because rotamers of
methyl groups by approx. 120 deg are mathematically
distinguishable but chemically identical. The trajectory
was sampled every 20 ps (10,001 snapshots). Time lag $\tau$
was set to 3 frames (60 ps). The value of perplexity was
set to 3.0 and Euclidean space was used to calculate
the distance matrix $\mathbf{D}$.

The value of lag time was chosen based on TICA results.
Similar calculations with lag time set to 1 to 12 steps show
that lag time set to 1-7 works well on a simple system such
as alanine dipeptide (see Supplementary Material).

\begin{figure}[h]
  \includegraphics{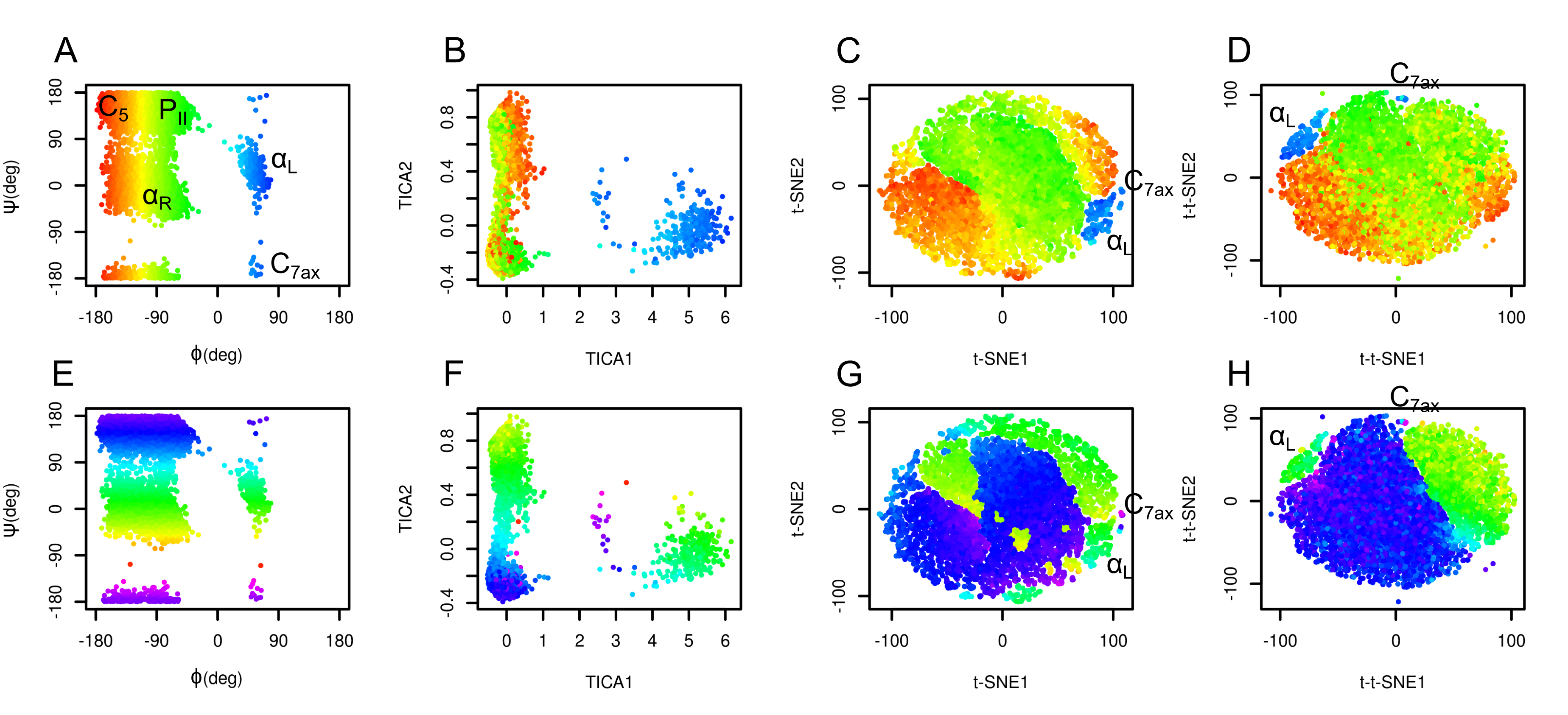}
  \caption{Time-lagged t-SNE (t-t-SNE) applied on 
  200 ns simulation of alanine dipeptide in water.
  Conformations sampled in
  the simulations were projected into the space of Ramachandran
  torsions $\phi$ and $\psi$ (A, E), TICA coordinates (B, F),
  t-SNE (C, G) and time-lagged t-SNE (D, H). Points are colored
  by Ramachandran torsion $\phi$ (A-D) and $\psi$ (E-H).}
  \label{fig:fig1}
\end{figure}

The results are depicted in Figure \ref{fig:fig1}. Plots in
the space of Ramachandran torsions show that all relevant
conformations of alanine dipeptide were sampled. Plots in
the space of TICA coordinates show that rotation around $\phi$
is the slowest and rotation around $\psi$ is the second slowest
motion in the studied system.

Plots in the space of t-SNE coordinates has a
yin-and-yang-like shape. These plots show a limitation
of conventional t-SNE, which is an improper resolution of
conformations. Namely, there are several yellow-green islands
in the blue area of the plot colored by $\phi$ values (G).

Time-lagged t-SNE (t-t-SNE) does not suffer this problem.
The blue area in the plot generated by time-lagged tSNE is
continuous and does not contain any islands of conformations
with positive $\phi$ values (H). This can be explained by the
fact that introduction of a time lag into t-SNE causes higher
separation of key conformations of alanine dipeptide.

One feature is common to the original t-SNE as well as our
time-lagged variant. This is the fact that t-SNE flattens
the distribution of points in the output space. This results
into almost uniform distribution of points in each minimum.

It is possible to calculate a histogram of some molecular
collective variable or collective variables and convert
it into a free energy surface. Most common interpretation
of such free energy surface is that deep minima correspond to
stable states wheres shallow minima correspond to unstable states.
This approach can be applied for conventional descriptors,
such as Ramachandran angles of alanine dipeptide. However,
due to flattening of distribution of points by t-SNE or by
time-lagged t-SNE such free energy surface is relatively flat.
Populations of different states can be calculated from areas
of free energy minima rather than from their depths.

\subsection{Trp-cage}
t-SNE and time-lagged t-SNE analysis was performed on
the trajectory of Trp-cage folding and unfolding sampled
every 20 ns (10,440 snapshots). Lag time was set to three
frames (60 ns). Perplexity was set to 10.0.

Similarly to alanine dipeptide, lag time was chosen based on TICA
analysis. Comparison of embeddings calculated for lag time set
to 1, 2, 3, 4, 5, 10, 15 and 20 (in number of frames) shows that
lag time 1 to 5 works well.

Initial analysis by time-lagged t-SNE resulted into
a circular plot with multiple points located outside
clusters of points on the edges of the circle.
This indicates that there are many points with high distances
$D_{t,t'}$. In order to eliminate these points we reduced
the number of eigenvectors $\mathbf{W^{\mathbf{Y}}}$ to top
50 eigenvectors (option \texttt{-maxpcs} in the code).

\begin{figure}[h]
  \includegraphics{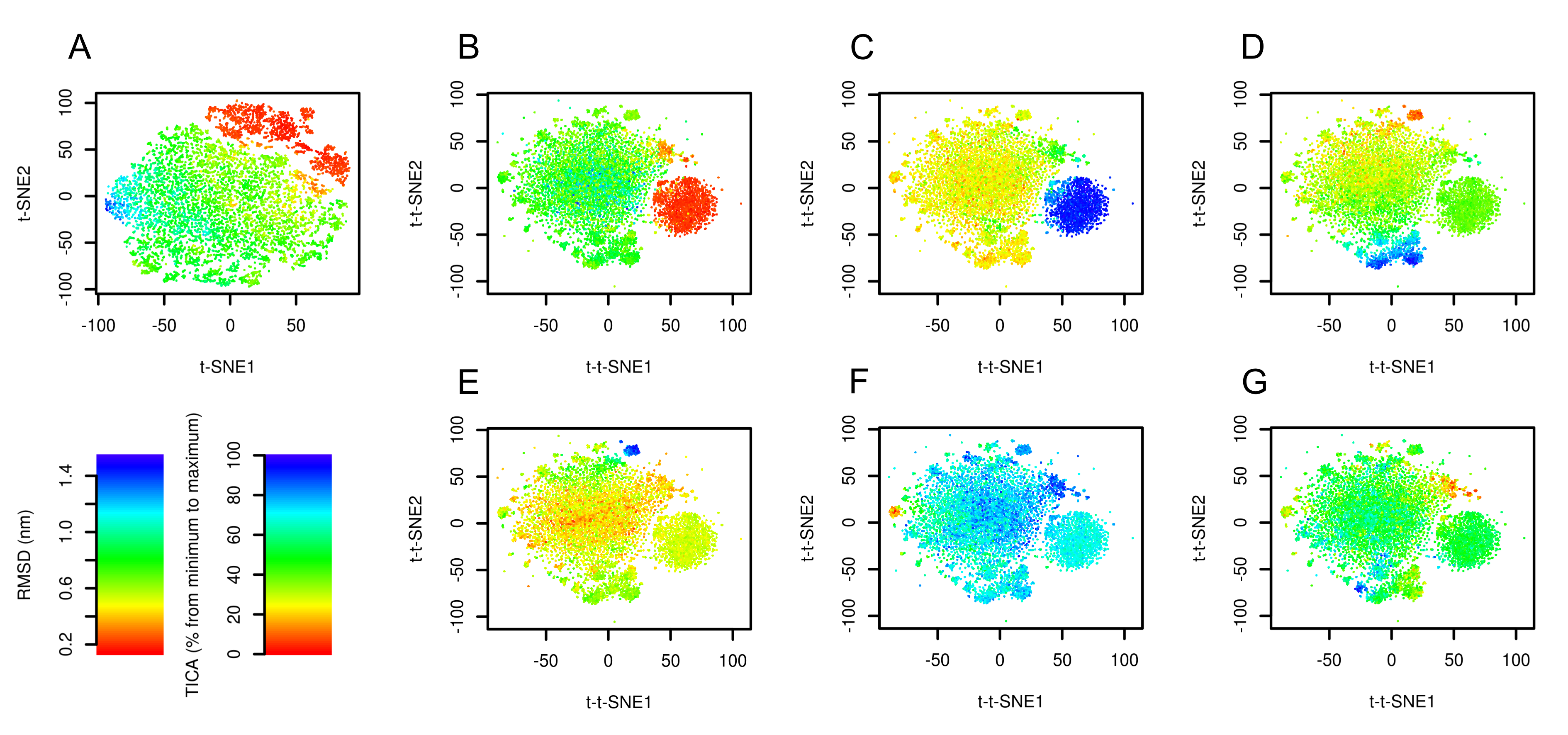}
  \caption{Time-lagged t-SNE (t-t-SNE) applied on 208.8 $\mu$s
  of Trp-cage folding and unfolding. The trajectory was
  analysed by t-SNE (A) and time-lagged t-SNE (B-G). Points
  are colored by RMSD from the native structure
  (A, B) and by the first (C), second (D),
  third (E), fourth (F) and fifth (G) TICA coordinate.}
  \label{fig:fig2}
\end{figure}

The results are depicted in Figure \ref{fig:fig2}.
Figure \ref{fig:fig2}A shows the trajectory analysed by conventional
t-SNE colored by RMSD from the native structure (PDB ID: 1l2y).
There is clear relationship between t-SNE coordinates, in
particular t-SNE1, and RMSD.
The native structure (in red) forms a cluster in top right corner
of the plot. Structures with high RMSD (in blue)
are characterized by lowest values of t-SNE1.

The trajectory analysed by time-lagged t-SNE colored by
RMSD is depicted in Figure \ref{fig:fig2}B. Similarly to
Figure \ref{fig:fig2}A the native structure forms a distinct
cluster. In contrast to the conventional t-SNE, structures
with high values of RMSD are scattered in the large cluster
in the center. This indicates that transitions between high-RMSD
structures are fast.

Figures \ref{fig:fig2}C-G show same plots colored by
TICA coordinates. The first TICA coordinate
(Figure \ref{fig:fig2}C) distinguishes folded and unfolded
structures. Plots colored by other TICA coordinates
(Figures \ref{fig:fig2}D-G) show usually a red or blue
clusters on edges of the plot. This shows that time-lagged
t-SNE captures slowest motions characterized by TICA, but 
more efficiently than TICA itself, because these motions can
be depicted in a single plot.

Figures \ref{fig:fig2}D, F and G show clusters with opposite
values of TICA coordinates (red vs. blue) on opposite sites
of plots. The interpretation is that conformations
with slow mutual transitions are located on distant locations
in the plot.

\begin{figure}
  \includegraphics{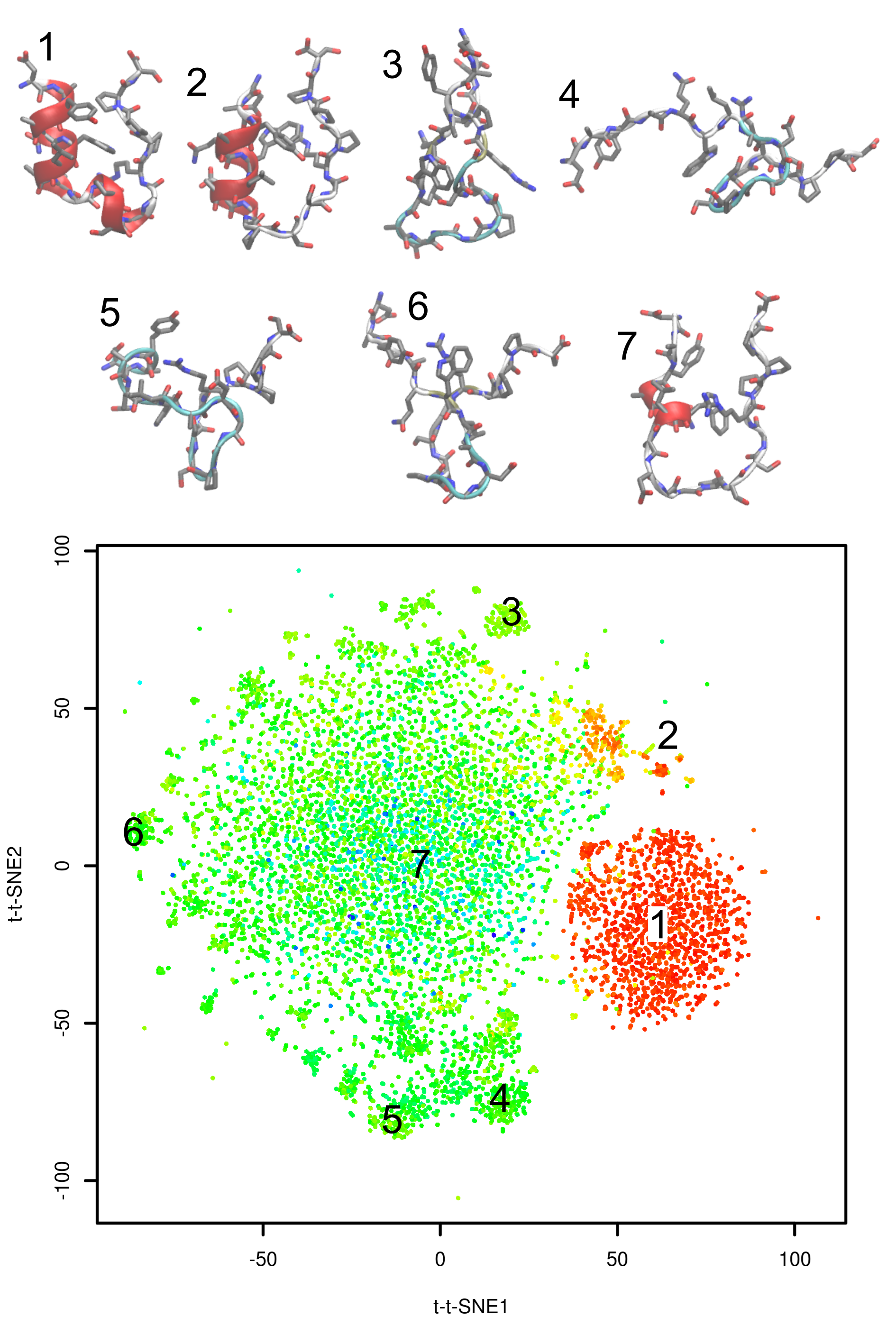}
  \centering
  \caption{Representative structures projected onto time-lagged
  t-SNE embeddings. Plot is colored by RMSD from the native structure
  (as in Fig. \ref{fig:fig2}B).}
  \label{fig:fig3}
\end{figure}

Figure \ref{fig:fig3} shows representative structures of Trp-cage
from the simulation trajectory projected onto time-lagged t-SNE
embeddings. Structure 1 is the native structure. Structure 2 is
a known near-native structure. Structures 3-6 were sampled from
clusters on peripheral areas of time-lagged t-SNE embeddings.
Finally, structure 7 was taken from the middle of the plot. Visual
inspection indicates that structures 3-6 may be kinetic traps
of Trp-cage folding, because these structures are characterized
by formation of numerous non-native hydrogen bonds and other
interactions. Also the near-native structure 2 is likely to be
a kinetic trap of Trp-cage folding.

\section{Discussion}
Choice of lag time for time-lagged t-SNE was driven by TICA
analysis. Values of 3 frames (60 ps, 0.03 \% of the whole
trajectory) for alanine dipeptide and 3 frames (60 ns,
0.029 \% of the whole trajectory) for Trp-cage led to visually
plausible low dimensional embeddings. This indicates, that
0.03 \% of trajectory size is good initial choice of lag time.

As an alternative to time-lagged t-SNE it is possible to
use time-lagged autoencoders recently reported by
Wehmeyer and Noé \cite{noe}. Autoencoders are feed-forward
neural networks with an hourglass-like architecture.
The input signal (atomic coordinates or other features)
from the input layer are transformed via
hidden layers into the central bottleneck layer. Next, the
signal from the bottleneck layer are transformed via another
hidden layers into the output layers. Parameters of the network
are trained to obtain agreement between the input and output
signal. The signal in the bottleneck layer represents
a non-linear low-dimensional representation of the input signal.
Unlike classical autoencoders, time-lagged autoencoders focus
on slowest motions, not on the most intensive motions \cite{noe}.

The clear advantage of autoencoders and their time-lagged
variant is the possibility to calculate low-dimensional
embeddings for a new out-of-sample structure.
Extensive testing of time-lagged autoencoders in the original 
article \cite{noe} was possible owing to this fact.
Time-lagged autoencoders can be trained on a training set 
and tested on a validation set, i.e. they can be evaluated
by cross-validation. Furthermore, they can be trained on
a small training set and then applied on a large set of
input data. This is efficient since the training part
is in general significantly more expensive than the calculation
of embeddings on out-of-sample structures. Time-lagged
autoencoders are useful for pre-processing of structural
data for building of Markov state models.

There are limited options for calculation of t-SNE low-dimensional
embeddings for out-of-sample structures. Therefore, t-SNE and
time-lagged t-SNE are not suitable for pre-processing of the
structural data. We see the advantage of time-lagged t-SNE
(similarly to t-SNE) in visualization.

Time-lagged t-SNE in the current implementation
also cannot be used as collective variables in simulations
using bias force or bias potential because these methods
require on-the-fly calculation of low-dimensional embeddings
and their derivatives with respect to atomic coordinates.
However, there are tools to approximate such low-dimensional
embeddings \cite{iso2,anncolvar}.

One of key features of t-SNE is that it can reconstruct 
proximities and not distances in the low-dimensional output
space. In time-lagged t-SNE this means that states separated
by low energy barriers are close to each other. States separated
by large energy barriers are far from each other, but time-lagged
t-SNE does not attempt to preserve their distances accurately. This means
that two close points in the time-lagged t-SNE plot can be connected
by an energetically favorable path.

Another key feature of t-SNE is perplexity and the fact that
t-SNE flattens the distribution of points in the output space.
This is useful for visualization. For this reason t-SNE
(as well as time-lagged t-SNE) must be used with caution as
a pre-processing for calculation of free energy surfaces
and for clustering. t-SNE can also create artificial cluster
when perplexity is not set properly.

\section{Acknowledgement}
This work was funded by COST action OpenMultiMed (CA15120,
Ministry of Education, Youth and Sports of the Czech Republic
LTC18074) and Czech National Infrastructure for Biological Data
(ELIXIR CZ, Ministry of Education, Youth and Sports of the Czech
Republic LM2015047). Authors would like to thank D. E. Shaw
Research for data used in this work.

\bibliographystyle{unsrt}

\end{document}


\maketitle

\section{Supplementary Material}

\begin{figure}
  \includegraphics{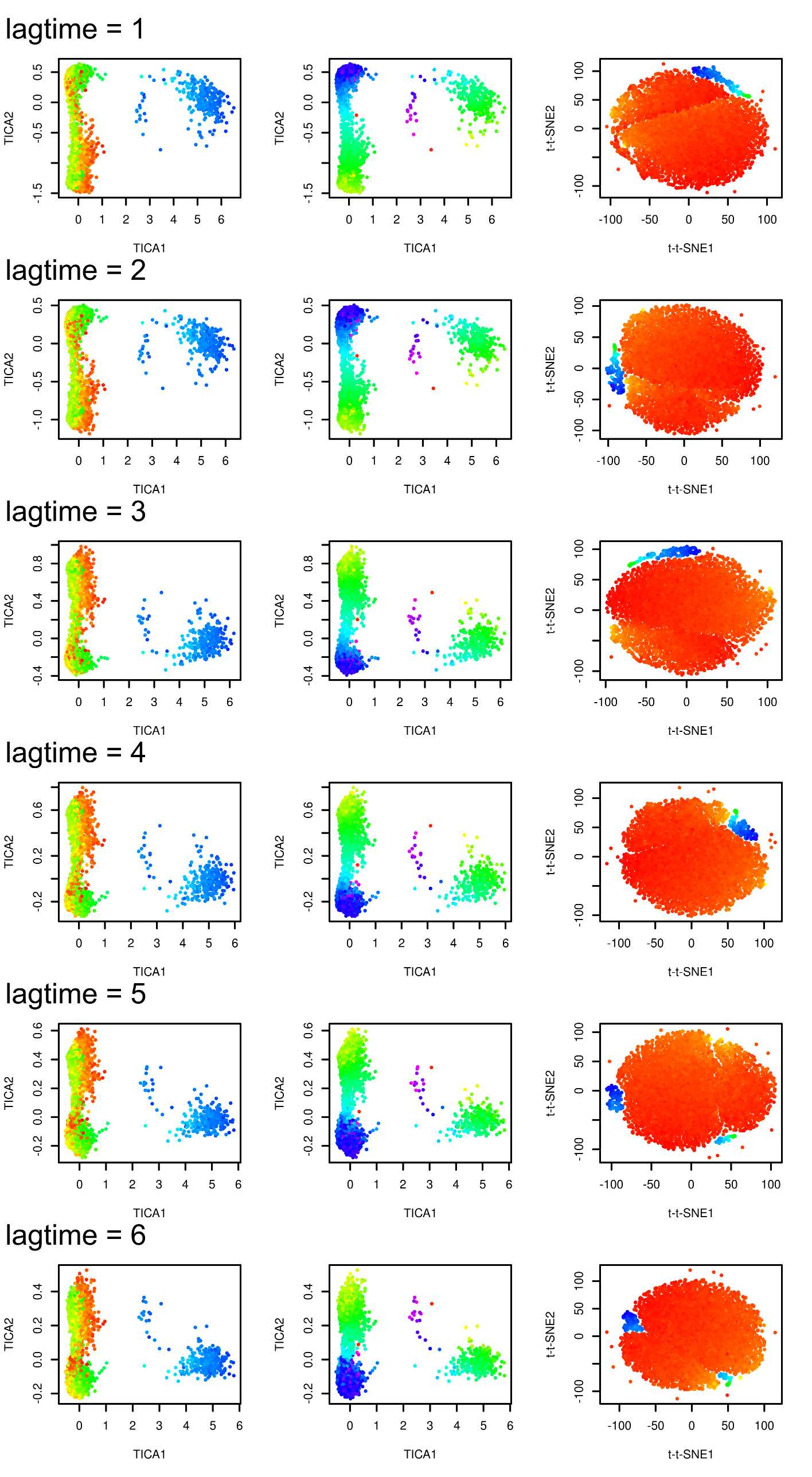}
  \centering
  \caption{Effect of lag time on time-lagged t-SNE of alanine
  dipeptide trajectory. TICA plots are colored by $\phi$ (left)
  and $\psi$ (centre, as in Fig. 1). Time-lagged t-SNE plots are colored by
  the first TICA coordinate (as in Fig. 1).}
  \label{fig:figS1}
\end{figure}

\begin{figure}
  \includegraphics{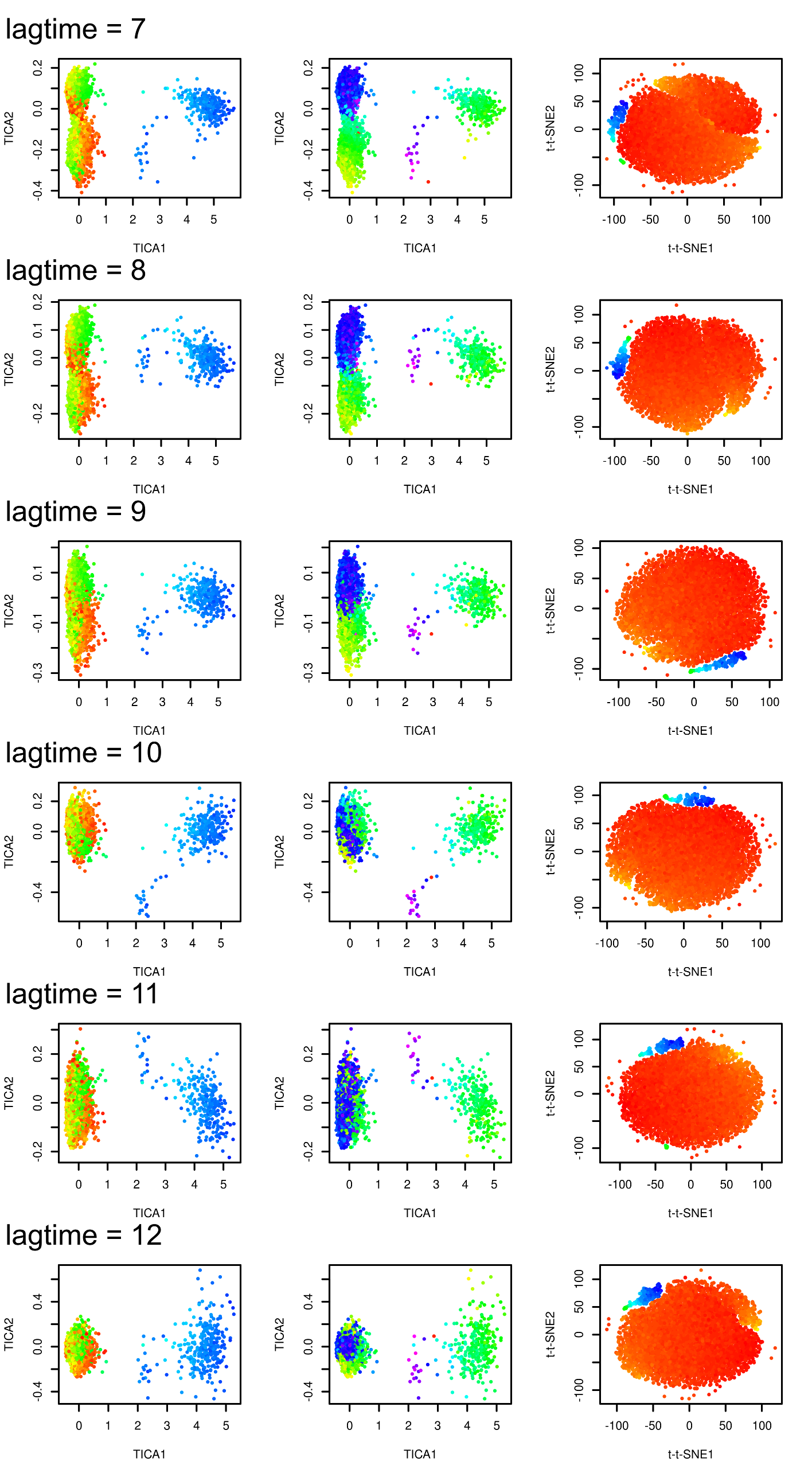}
  \centering
  \caption{Effect of lag time on time-lagged t-SNE of alanine
  dipeptide trajectory. TICA plots are colored by $\phi$ (left)
  and $\psi$ (centre, as in Fig. 1). Time-lagged t-SNE plots are colored by
  the first TICA coordinate (as in Fig. 1).}
  \label{fig:figS2}
\end{figure}

\begin{figure}
  \includegraphics{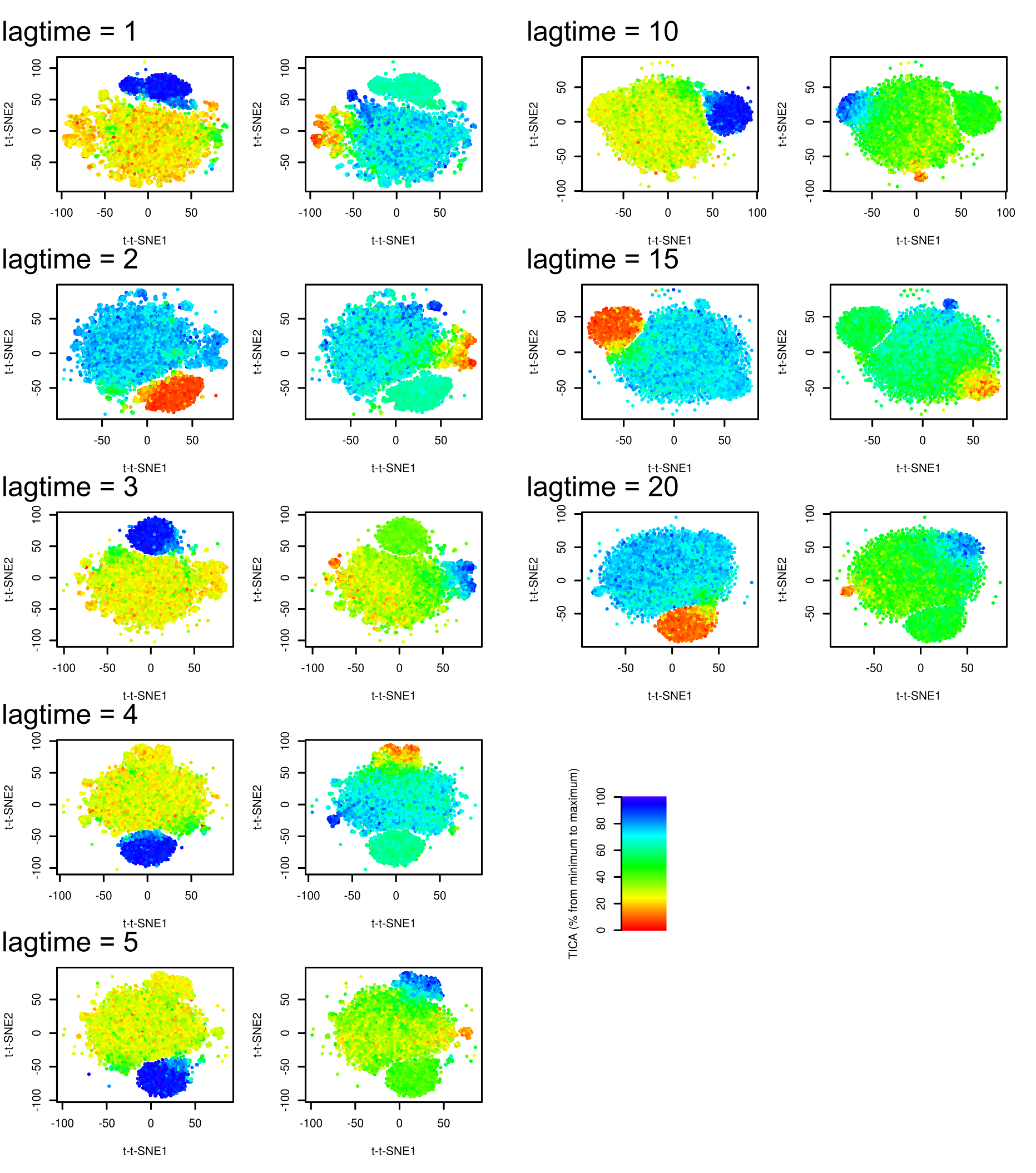}
  \centering
  \caption{Effect of lag time on time-lagged t-SNE of Trp-cage
  folding trajectory. Plots are colored by the first and the second
  TICA coordinate (left and right, respectively).}
  \label{fig:figS3}
\end{figure}

Supplementary figures Fig. \ref{fig:figS1}, Fig. \ref{fig:figS2} and
Fig. \ref{fig:figS3}.

